%% file: HIN-22-003_temp.tex
\documentclass[11pt,twoside,a4paper,cmspaper,final,collab]{cms-tdr}
\def\svnVersion{223caf0-D}\def\svnDate{2026/06/15}\def\cmsCernNoTag{CERN-EP-2026-119}\def\cmsCernDate{\today}\def\cmsMessage{Submitted to Physics Letters B}
\begin{document}\cmsNoteHeader{HIN-22-003}

\newlength\cmsFigWidth
\ifthenelse{\boolean{cms@external}}{
  \setlength\cmsFigWidth{0.85\columnwidth}
  \providecommand{\cmsLeft}{upper\xspace}
  \providecommand{\cmsLLeft}{Upper\xspace}
  \providecommand{\cmsRight}{lower\xspace}
  \providecommand{\cmsRRight}{Lower\xspace}
  \providecommand{\cmsTable}[1]{\resizebox{\columnwidth}{!}{#1}}
}{
  \setlength\cmsFigWidth{0.4\textwidth}
  \providecommand{\cmsLeft}{left\xspace}
  \providecommand{\cmsLLeft}{Left\xspace}
  \providecommand{\cmsRight}{right\xspace}
  \providecommand{\cmsRRight}{Right\xspace}
  \providecommand{\cmsTable}[1]{#1}
}

\newcommand{\sqrtsNNeight}{\ensuremath{\sqrt{\smash[b]{s_{_{\mathrm{NN}}}}}=8.16\TeV}\xspace}
\newcommand{\rap}{\ensuremath{y_{\mathrm{CM}}}\xspace}
\newcommand{\ylab}{\ensuremath{y_{\mathrm{lab}}}\xspace}
\newcommand{\ylabFlip}{\ensuremath{y_{\mathrm{lab, p}}}\xspace}
\newcommand{\ptJ}{\ensuremath{p_{\mathrm{T}}(\JPsi)}\xspace}
\newcommand{\ylabFlipJ}{\ensuremath{y_{\mathrm{lab, p}}(\JPsi)}\xspace}
\newcommand{\pizero}{\PGpz\xspace}
\newcommand{\Jzerozero}{\ensuremath{J_{z}(\chicone)=0, J_{z}(\chictwo)=0}\xspace}

\newcommand{\pp}{\ensuremath{\Pp\Pp}\xspace}
\newcommand{\pPb}{\ensuremath{\Pp\mathrm{Pb}}\xspace}
\newcommand{\Pbp}{\ensuremath{\mathrm{Pb}\Pp}\xspace}

\newcommand{\Ntracks}{\ensuremath{N_{\text{tracks}}}\xspace}

\newcommand{\EPOS}{\textsc{epos}\xspace}
\newcommand{\photon}{\PGg\xspace}
\newcommand{\chic}{\PGcc}
\newcommand{\chib}{\PGcb}
\newcommand{\chiczero}{\HepParticle{\PGc}{c0}{}{}\xspace}
\newcommand{\chicone}{\HepParticle{\PGc}{c1}{}{}\xspace}
\newcommand{\chictwo}{\HepParticle{\PGc}{c2}{}{}\xspace}
\newcommand{\chictoJg}{\ensuremath{\PGcc \to \JPsi \PGg}\xspace}
\newcommand{\chictoJgfull}{\ensuremath{\PGcc \to \JPsi \, \PGg \to \PGmp\,\PGmm + \Pep\,\Pem}\xspace}
\newcommand{\mumu}{\ensuremath{\PGmp\PGmm}\xspace}
\newcommand{\ee}{\ensuremath{\Pep\Pem}\xspace}
\newcommand{\psitwo}{\PGyP{2S}}
\newcommand{\chictoJratio}{\chic-to-\JPsi ratio\xspace}
\newcommand{\chictwooneratio}{\chictwo-to-\chicone ratio\xspace}

\newcommand{\chica}{\ensuremath{\chicone \to \JPsi \, \photon}\xspace}
\newcommand{\chicb}{\ensuremath{\chictwo \to \JPsi \, \photon}\xspace}
\newcommand{\chicc}{\ensuremath{\chiczero \to \JPsi \, \photon}\xspace}
\newcommand{\mumugamma}{\ensuremath{\PGmp\PGmm \, \PGg}\xspace}

\cmsNoteHeader{HIN-22-003}

\title{Study of \texorpdfstring{\chic}{chi c} production in \pPb collisions at \texorpdfstring{$\sqrt{s_{\mathrm{NN}}}=8.16\TeV$}{sqrt(sNN) = 8.16 TeV}}

\author[cern]{The CMS Collaboration}

\date{\today}

\abstract{Production of prompt P-wave charmonium states $\chicone$(1P)  and $\chictwo$(1P) is studied in proton-lead (pPb) collisions at a center-of-mass energy per nucleon pair of $\ensuremath{\sqrt{\smash[b]{s_{_{\mathrm{NN}}}}}} = 8.16$ TeV. The analysis is based on data corresponding to an integrated luminosity of 175\nbinv collected by the CMS experiment at the CERN LHC. The $\chic$ states are measured via their decay $\chic \to \JPsi \, \PGg$. The \JPsi meson is reconstructed via its decay to a muon pair, while the photon is reconstructed through its conversion to an electron-positron pair. The ratio of production cross sections times branching fractions for the two charmonium states, $[\sigma(\chictwo) \mathcal{B}(\chictwo \to \JPsi \, \PGg)] / [\sigma(\chicone) \mathcal{B}(\chicone \to \JPsi \, \PGg)]$, is reported in the rapidity range $\abs{y(\JPsi)} < 2.4$ for the transverse momentum range $6.5 < p_{\mathrm{T}}(\JPsi) < 30$ GeV. The $\chictwo$-to-$\chicone$ ratio is found to be independent of event charged-particle multiplicity, as well as of the rapidity and $p_\mathrm{T}$ of the \JPsi. The consistency of this ratio with proton-proton measurements at $\ensuremath{\sqrt{s}}\xspace = 7$ TeV indicates an absence of strong relative modification for $\chic$ states in pPb collisions, in contrast to the behavior observed in the \PGyP{2S}-to-\JPsi ratio.}

\hypersetup{%
pdfauthor={CMS Collaboration},%
pdftitle={Chic production in pPb collisions at CMS},%
pdfsubject={CMS, chic},%
pdfkeywords={CMS, chic, heavy-ion collisions}}

\maketitle

\section{Introduction}

The studies of quarkonia production in proton-nucleus collisions play a key role in understanding the interplay between initial- and final-state cold nuclear matter effects~\cite{bib:QuarkoniaAtLHCEraOverview}. In processes that produce charmonia, effects of initial-state parton kinematics are expected to be similar in all states.
In addition to these initial-state effects, heavy quarkonia can be also subject to coherent energy loss of the evolving partonic system \cite{Arleo2013} and to breakup by comoving particles \cite{Ferreiro:2014bia, Capella:1996va}. 
In nucleus-nucleus collisions, quarkonium production can be further modified by the formation of a deconfined quark-gluon plasma (QGP), which provides a hot nuclear medium. Proton-nucleus measurements are therefore essential for isolating cold nuclear matter effects and for providing a baseline for interpreting quarkonium modifications in nucleus-nucleus collisions.

Earlier studies of particle production in p-nucleus collisions are reviewed in Ref.~\cite{HIN-23-011}. These works have typically focused on the quarkonia S-wave states, such as the \JPsi, \psitwo, and \PgU mesons \cite{LHCb:2017ygo, PHENIX:2016vmz, CMS:2022wfi, ATLAS:2017prf,ALICE:psi2S}.
Previous measurements of charmonium suppression in proton-nucleus collisions, in particular those pertaining to the \psitwo meson, have shown that this excited state has a different suppression behavior compared to the ground state \JPsi meson \cite{LHCb:2017ygo, PHENIX:2016vmz, ATLAS:2017prf, bib:Psi2SJpsiRdAuPHENIX}.
A qualitatively similar suppression pattern between excited and ground states has also been observed in the bottomonium states \cite{CMS:2022wfi, ATLAS:2017prf, LHCb:2018psc, ALICE:psi2S}.
These differences cannot be explained by initial-state effects alone, because those effects modify all quarkonia states in a similar way. In particular, a trend of increasing relative suppression of the \psitwo meson with increasing charged-particle multiplicity (or related variables) is observed~\cite{PHENIX:2016vmz, ATLAS:2017prf, bib:Psi2SJpsiRdAuPHENIX}.
By looking at distinct members of the charmonium family with different binding energies, sizes, angular momenta, and interaction cross sections with comovers~\cite{Ferreiro:2014bia}, one can investigate how final-state interactions drive the relative nuclear modification of different quarkonium states.

In this paper, we extend the measurements to the much less explored P-wave states, namely to the \chic(1P) states, referred to as \chic for the remainder of the paper, where \chic is used as a generic term for both \chicone(1P) and \chictwo(1P).
The binding energy and the size of the \chic states are between those of the \JPsi and \psitwo mesons \cite{Satz2006}. Studies of the \chic states can help to obtain a more complete picture of the underlying mechanisms that lead to modification of heavy quarkonium states in both cold and hot nuclear matter~\cite{MatsuiSatz, Ferreiro:2014bia, bib:SequentialSuppression}.

Recently, the LHCb Collaboration reported the first measurement of the cross section ratio of \chictwo-to-\chicone mesons, $\sigma(\chictwo)/\sigma(\chicone)$~\cite{bib:chictochicpPbLHCb}, and of \chic-to-\JPsi mesons, $\sigma(\chic \to \JPsi)/\sigma(\JPsi)$~\cite{bib:chictojpsipPbLHCb}, using promptly produced mesons from proton-lead (\pPb) collisions in the far-forward region ($1.5$ $<$ \rap $<$ $4.0$ , $-5.0$ $<$ \rap $<$ $-2.5$) at Large Hadron Collider (LHC) energies. The results indicate that the \chic contribution to prompt \JPsi production in \pPb collisions is compatible with expectations from pp measurements within uncertainties.
Additionally, several \chic measurements have been reported in \pp collisions, which serve as a baseline for studying the modifications in heavy ion collisions.
For example, measurements of the ratio of \chic to \JPsi mesons in pp collisions were reported in the midrapidity region by the ATLAS Collaboration~\cite{bib:chicATLAS2014}, and in the forward rapidity region by LHCb~\cite{bib:chictoJratioLHCb}. Measurements of $\sigma(\chictwo)/\sigma(\chicone)$ were also reported by CMS ~\cite{bib:chictochicratioCMS2012}, LHCb~\cite{bib:chictochicratioLHCb}, and ATLAS~\cite{bib:chicATLAS2014}.

We measure the \chictwooneratio: the ratio of production cross sections times branching ratios of the \chicone and \chictwo states. The \chic candidates are reconstructed through the $\chic \to \JPsi \, \PGg$ radiative decays. The branching fraction $\mathcal{B}(\chicc)$ is only 1.4\%, compared to $\mathcal{B}(\chica) = 34\%$ and $\mathcal{B}(\chicb) = 19\%$, and therefore the \chiczero state is undetectable in this analysis.
The \JPsi meson is reconstructed through the $\cPJgy \to \mumu$ decay channel and the photon via the conversion into an \ee pair in the detector material.

The \chictwo-to-\chicone ratio is defined as
\begin{equation}
\frac{\sigma\!\left(\pPb \to \chictwo X\right)\,\mathcal{B}(\chicb)}{\sigma\!\left(\pPb \to \chicone X\right)\,\mathcal{B}(\chica)}.
\label{eq:chictwooneratio}
\end{equation}
The branching fraction $\mathcal{B}(\JPsi \to \mumu)$ appears in both the numerator and the denominator, and therefore cancels out.
The \chic and \JPsi mesons produced from b-hadron decays are removed in the analysis, retaining only prompt \chic mesons. 
The measurement is performed using the data sample of \pPb collisions at a center-of-mass energy per nucleon pair of \sqrtsNNeight, corresponding to an integrated luminosity of $175\pm 7$\unit{nb$^{-1}$}, collected by the CMS experiment in 2016 \cite{bib:CMSlumipPb, bib:CMSlumipp2016}. The results of this analysis are also provided as a HEPData record ~\cite{HEPData}.

\section{The CMS detector}
\label{sec:CMSdetector}

A detailed description of the CMS detector can be found in Ref.~\cite{CMS_2008}. The inner tracker, composed of a pixel detector and a silicon-strip detector and immersed in a 3.8\unit{T} solenoidal magnetic field, is located at the center of the detector. It measures charged-particle trajectories in the pseudorapidity range $\abs{\eta}<2.51$.

Outside the solenoid, there are up to four layers of gas-ionization muon detectors sandwiched between the elements of the steel return yoke of the magnet. The muon system covers the region $\abs{\eta}<2.4$ and performs three main tasks: triggering on, identifying, and improving the momentum measurement and charge determination of muons with high transverse momentum \pt. Drift tube chambers and cathode strip chambers detect muons in the regions of $\abs{\eta}<1.2$ and $0.9<\abs{\eta}<2.4$, respectively. These two detector technologies are complemented by a system of resistive plate chambers covering the range of $\abs{\eta}<1.6$. The use of these different technologies defines three regions in the detector, referred to as barrel ($\abs{\eta}<0.9$), overlap ($0.9<\abs{\eta}<1.2$), and endcap ($1.2<\abs{\eta}<2.4$). Muon energy deposits in the electromagnetic calorimeter (ECAL), hadronic calorimeter, and outer hadronic calorimeter are also used for muon identification.

Events of interest are selected using a two-tiered trigger system~\cite{Khachatryan:2016bia}. The first level, composed of custom hardware processors, uses information from the calorimeters and muon detectors in order to select events at a rate of approximately $100\unit{kHz}$ within a fixed time interval of $\approx$ $4\mus$. The second level, the high-level trigger (HLT), consists of a farm of processors running a version of the full event reconstruction software, including the inner tracker, optimized for fast processing. The HLT reduces the event rate to around 1.6\unit{kHz} before data storage. 
The data set for this analysis was collected using a trigger requiring two muon candidates at the first level, with no additional HLT requirement.

The final-state particles are reconstructed with a particle-flow algorithm \cite{CMS_Particle_flow}, which combines information from all subdetectors to reconstruct individual particle candidates.
These are classified as muons, electrons, photons, charged and neutral hadrons. Events are required to have at least one reconstructed primary vertex (PV), formed from two or more associated tracks. The PV must be located within 25 cm of the detector center along the z-axis and within a 2 cm radius in the x-y plane \cite{Chatrchyan:2014fea}.

Photon conversions are characterized by an oppositely charged track pair originating from a common vertex, with an \ee invariant mass consistent with zero.
The electron tracking algorithm is described in Ref.~\cite{Chatrchyan:2014fea}.
A correction factor is applied to photon reconstruction, accounting for the probability of \ee conversion occurring and the efficiency of reconstructing the converted pair by the tracking algorithm.
The conversion probability is low, ranging from 1-16\%, depending on the phase space, because the inner tracking layers being optimized for a low material budget.
Nevertheless, this analysis relies on conversions because the momentum resolution for the low-momentum photons from the \chictoJgfull decay is superior to that obtained from the ECAL \cite{bib:PhotonPerf}.

\section{Data and simulated samples}
\label{sec:DataMC}

The data taking for \pPb collisions was divided into two periods, with the direction of each colliding beam reversed after the first period.
The beam-direction reversal is employed to mitigate and control potential detector asymmetries with respect to the rapidity, especially considering that the center-of-mass (CM) in \pPb collisions is not stationary but shifted towards the proton beam direction (the p beam energy is 6.5 \TeV and the Pb beam energy is 2.56 \TeV), with a rapidity shift of $\rap = 0.465$.
The integrated luminosity collected by the trigger is $63 \pm 2$\nbinv and $112 \pm 5$\nbinv for the Pbp and pPb periods, respectively. The beam directions are defined with respect to the CMS coordinate system: for pPb (Pbp) collisions, the proton (lead nucleus) is defined to move along the +$z$ direction~\cite{bib:CMSlumipPb, bib:CMSlumipp2016}. In this dataset, the average number of collisions per bunch crossing varied between 0.1 and 0.25. There is a non-negligible probability of multiple collisions occurring in the same bunch crossing, referred to as pileup. To reduce the impact of such pileup interactions, the variable \Ntracks is defined as the number of tracks originating from the PV associated to the dimuon and passing the following quality criteria: $\pt > 0.4$ GeV, $\abs{\eta} < 2.4$, relative \pt resolution better than 0.1, and 3$\sigma$ compatibility with the PV in $xy$- and $z$-directions.

Monte Carlo (MC) simulations of \chictoJg signals  are used for efficiency and acceptance corrections, as well as for modeling the invariant mass probability density functions.
They are generated with {\PYTHIA} (v8.212)~\cite{bib:Pythia} for production and hadronization processes, and with \EVTGEN (v1.3.0) for simulation of the \chic meson decays ~\cite{bib:EvtGen}. The quantum-electrodynamics final-state radiation is modeled by \PHOTOS (v3.56)~\cite{bib:Photos} at leading order. Each \PYTHIA event is embedded in a \pPb collision event that is simulated with \EPOS \textsc{lhc} (v3400)~\cite{bib:EPOS} to reproduce the underlying backgrounds. The generated events are then processed through a detailed simulation of the CMS detector, based on the \GEANTfour package~\cite{bib:Geant4}, using the same trigger and reconstruction algorithms as were used for the collision data. 
Separate MC samples are used for each beam direction and are weighted according to the recorded luminosities for the \pPb or \Pbp periods.  The MC samples are also weighted to match the data distributions that are most relevant for efficiency corrections. 
These include \ptJ from the \chictoJg decays, and \Ntracks.

\section{Signal extraction}
\label{sec:SignalExtraction}

\subsection{Particle selection}
\label{sec:Selection}

The \chic candidates are reconstructed via the \chictoJgfull decay chain. Since \chic decays promptly, the \JPsi is assumed to originate from the PV for the purpose of the \chic reconstruction.
The reconstruction starts by selecting muons with the \textit{soft muon} identification, which requires that the muon is reconstructed as a tracker muon with at least one tightly matched muon segment and that it has a high-purity track with sufficient hits in pixel and strip layers~\cite{bib:CMS_MUO-10-002, bib:CMS_MUO-16-001}. The muons are required to be in the acceptance region defined in Table~\ref{tab:muonAcc}. These thresholds are chosen to ensure stable double-muon trigger and reconstruction efficiencies across the detector acceptance. Additionally, each muon is required to meet the trigger condition defined in Section~\ref{sec:CMSdetector} to ensure that the dimuon pair is responsible for triggering the event.

\begin{table}[ht!]

	\centering
  \renewcommand{\arraystretch}{1.0} 
  \topcaption{Muon $\abs{\eta}$ and \pt acceptance used in the analysis.}
 
  \begin{tabular}{rl}
  $\abs{\eta}$ range & $\pt$ range $[\mathrm{GeV}]$ \\ \hline
  $0 < \abs{\eta} < 0.3$   & $\pt>3.4$ \\
  $0.3 < \abs{\eta} < 1.1$ & $\pt>3.3$ \\
  $1.1 < \abs{\eta} < 2.1$ & $\pt>5.5-2.0\abs{\eta}$ \\
  $2.1 < \abs{\eta} < 2.4$ & $\pt>1.3$ \\
  \end{tabular}
  \label{tab:muonAcc}
\end{table}

Next, the \JPsi candidates are formed by combining all possible muon pairs from the selected muons within a single event. We require that the muons of a pair have opposite charges and that the \JPsi candidates satisfy  $\abs{\ylab(\JPsi)}<2.4$ and $ 6.5 < \pt(\JPsi) < 30\GeV$, where \ylab is the lab-frame rapidity. 
A dimuon vertex fit is performed on the muon pair with a requirement that the probability of the muon tracks originating from a common vertex must be greater than 1\%.
Contributions from non-prompt decays of $B$ mesons are reduced by applying a requirement on the two-dimensional proper decay length (hereafter referred to as the lifetime), optimized using prompt \chic simulation studies: $c\tau=cL_{xy}(M/\pt)$, where $L_{xy}$ is the distance between the \JPsi meson decay point and the PV, projected onto the $xy$-plane. We determined and applied a proper decay length selection derived from prompt \chic simulation ($c\tau/\sigma(c\tau) < 3$, $\sigma(c\tau)\approx 39\mum$).

The photon conversion selection begins with the removal of duplicate conversion candidates. These are candidates where one of the tracks is split in the tracker, resulting in a second candidate with almost identical kinematics. These duplicate conversions are removed by keeping only the one that uses the track with the larger number of valid hits. For the photon, after the initial photon conversion candidates are found, we apply the following quality criteria:

\begin{itemize}

\item Compatible inner hits:  The two tracks from a candidate conversion  must have one of the two innermost hits in the same tracker layer.  This is to reduce the contribution of misidentified photon conversions that originate from low-momentum displaced tracks that are spuriously propagated backwards.
\item $\Delta z$ to the closest PV:	The difference between the $z$ coordinate where the reconstructed conversion candidate would cross the beam line (by propagating its momentum backwards) and the $z$ coordinate of the PV must satisfy $\Delta z<10\,$cm.
\item Distance of closest approach (DCA): The conversion distance of closest approach in the $x$-$y$ (transverse) plane must satisfy the condition $DCA > -10\,$ cm. The $DCA$ variable is defined as the distance between the centers of the two circles ($O_1$ and $O_2$) made by projecting the track helices onto the transverse plane, minus the sum of their radii ($R_1$ and $R_2$), \ie, $DCA = \abs{O_1 - O_2} - (R_1+R_2)$. Ideally, the tracks would point exactly to the common vertex, thus $DCA$ being equal to 0. The experimental uncertainty in the $DCA$ measurements is about $0.2\,$cm. The value can be negative if the two circles overlap. The selection of $DCA > -10\,$ cm is applied in order to reject poor quality conversion vertices. 
\end{itemize}

The selection criteria are based on those used in the previous CMS $\chi_c$ analysis~\cite{bib:chictochicratioCMS2012} and were reoptimized for the present pPb data set by comparing several cut-based selections with boosted-decision-tree-based alternatives. The cut-based selection used in this analysis was chosen as it gives a signal efficiency close to the best boosted decision tree performance while avoiding additional complexity in the selection. Finally, the \chictoJg candidates are formed by combining the \JPsi candidate with a converted photon. The dimuon candidate mass is required to be between 2.9-3.25\GeV, in order to reduce the contribution from the combinatorial background. The \chic candidate mass is calculated using the world average of experimentally determined values of the \JPsi mass \cite{bib:PDG2020} instead of the invariant mass of the dimuon pair, $m(\chic)= m(\mumugamma) - m(\mumu) + 3.097$\GeV. This improves the \chic meson mass resolution by reducing the propagation of event-by-event fluctuations in the reconstructed dimuon mass to the \chic candidate mass.

\subsection{Yield extraction}
\label{sec:Fitting}

After the  \JPsi and \chic candidate selection, we fit the $m(\chic)$ distribution. 
The \chictwooneratio (Eq.(~\ref{eq:chictwooneratio})) is reported as a function of the kinematic variables of the \JPsi meson. This convention is adopted to allow comparisons to prior results that used the same methodology. Based on the two-body decay kinematics, the \JPsi meson carries most of the parent \chic meson momentum, while the photon is soft. Therefore, the \chic and the descendant \JPsi meson kinematic distributions are similar. In order to combine the \pPb/\Pbp data sets, the p-going direction is defined as positive, meaning that the sign of the rapidity for the \Pbp portion of the run is reversed in the CMS coordinate system. The combined rapidity variable is denoted as \ylabFlip in this paper.

\subsubsection{The \texorpdfstring{\chic}{Chi c} candidate fits}
\label{sec:Fitting_chic}

Fitting the \chic signal is challenging because of the low signal-to-background ratio, and steps must be taken in order to obtain good-quality fits. Our chosen signal model consists of a double-sided crystal-ball function (DCB), which is an extension of the CB function with power-law tails on both sides. 
The full definition of the DCB function is

\begin{equation}
\begin{array}{@{}l@{}}
\mathrm{DCB}(m; \mu, \sigma, \alpha_{\textrm{L}}, n_{\textrm{L}}, \alpha_{\textrm{H}}, n_{\textrm{H}}) = \\[0.8ex]
\begin{cases}
      \re^{-0.5 \, t^{2}} & \quad \mbox{if $-\alpha_{\textrm{L}} \le t \le \alpha_{\textrm{H}}$,} \\
      \re^{-0.5 \, \alpha_{\textrm{L}}^{2} } \left[\frac{\alpha_{\textrm{L}}}{n_{\textrm{L}}} \left(\frac{n_{\textrm{L}}}{\alpha_{\textrm{L}}} - \alpha_{\textrm{L}} - t \right)\right]^{-n_{\textrm{L}}}  & \quad \mbox {if $t < -\alpha_{\textrm{L}}$,} \\
      \re^{-0.5 \, \alpha_{\textrm{H}}^{2} } \left[\frac{\alpha_{\textrm{H}}}{n_{\textrm{H}}} \left(\frac{n_{\textrm{H}}}{\alpha_{\textrm{H}}} - \alpha_{\textrm{H}} + t \right)\right]^{-n_{\textrm{H}}}  & \quad \mbox {if $t > \alpha_{\textrm{H}}$,} \\
    \end{cases}
\end{array}
\label{eq:DCB}
\end{equation}

where $t = (m - \mu)/\sigma$. The subscripts $\textrm{L}$ and $\textrm{H}$ refer to the low- and high-mass tail parameters, respectively. The {\PYTHIA} MC sample was used to determine the six DCB parameters for the signal shape of each  of the two \chic states. A \textit{threshold} function is used for the background, defined as

\begin{equation}
\mathrm{BKG}(m; \alpha_1, \beta_1) = (m-q_0)^{ \alpha_1} \re^{(m-q_0) \beta_{1}},
\label{eq:Threshold}
\end{equation}

where $\alpha_{1}$ and $\beta_{1}$ are free parameters, and $q_0$ is fixed to 3.2\GeV corresponding to the sum of the \JPsi mass and the minimal \pt requirement of converted photon. The threshold function was used in previous CMS \chic and \chib analyses in \pp collisions~\cite{bib:chictochicratioCMS2012, bib:Khachatryan:2014ofa, bib:Sirunyan:2018chib3p}, and it was found to be applicable in the \pPb collisions as well. It describes the background while having a reasonably low number of parameters.
The overall fit function in the \chic invariant mass region consists of two DCB functions, one for the \chicone peak and the other for the \chictwo peak, added to the background. The \(\chictwo/\chicone\) yield ratio is included as a free parameter in the fit, so its statistical uncertainty is obtained directly from the fit result and includes correlations with the other floating fit parameters.

\begin{figure}[hbt!]
  \centering
  \includegraphics[width=0.48\textwidth]{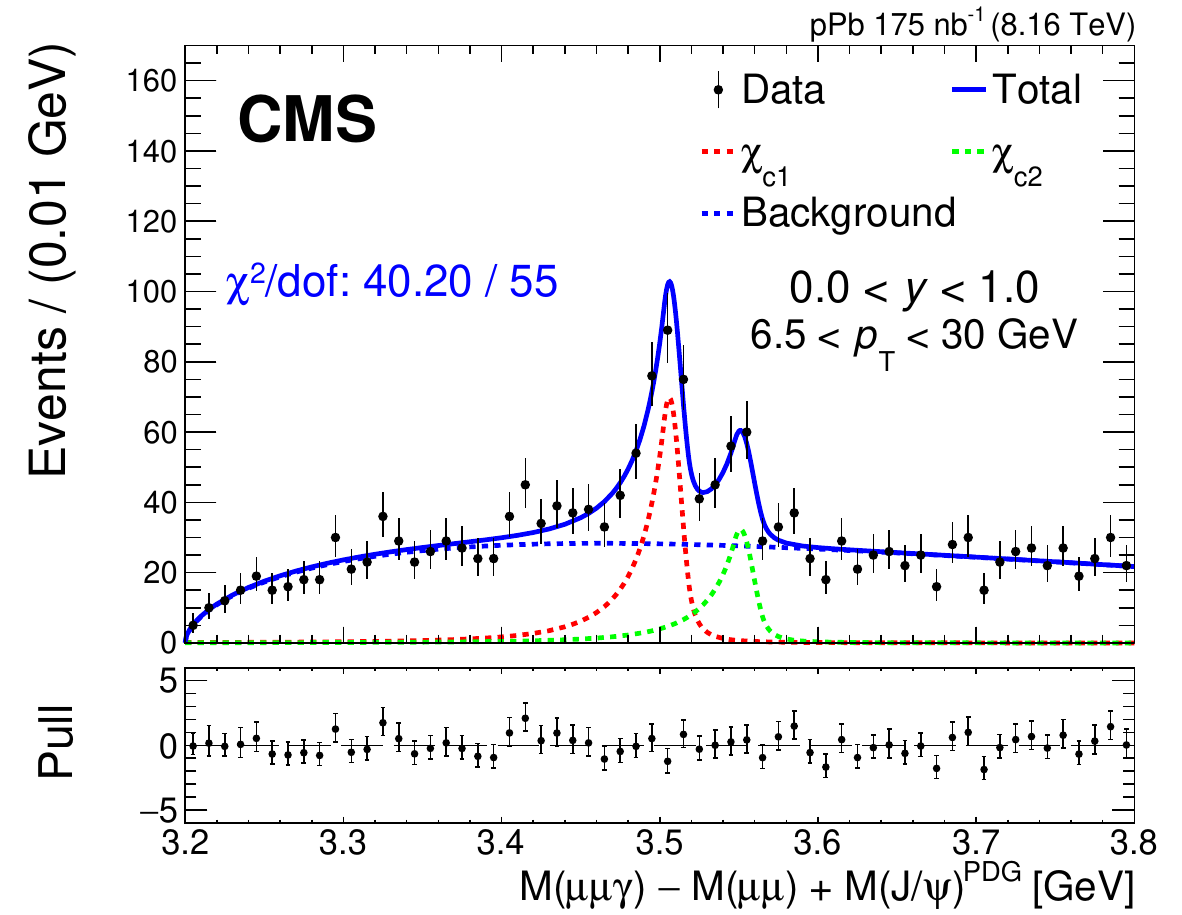}
  \includegraphics[width=0.48\textwidth]{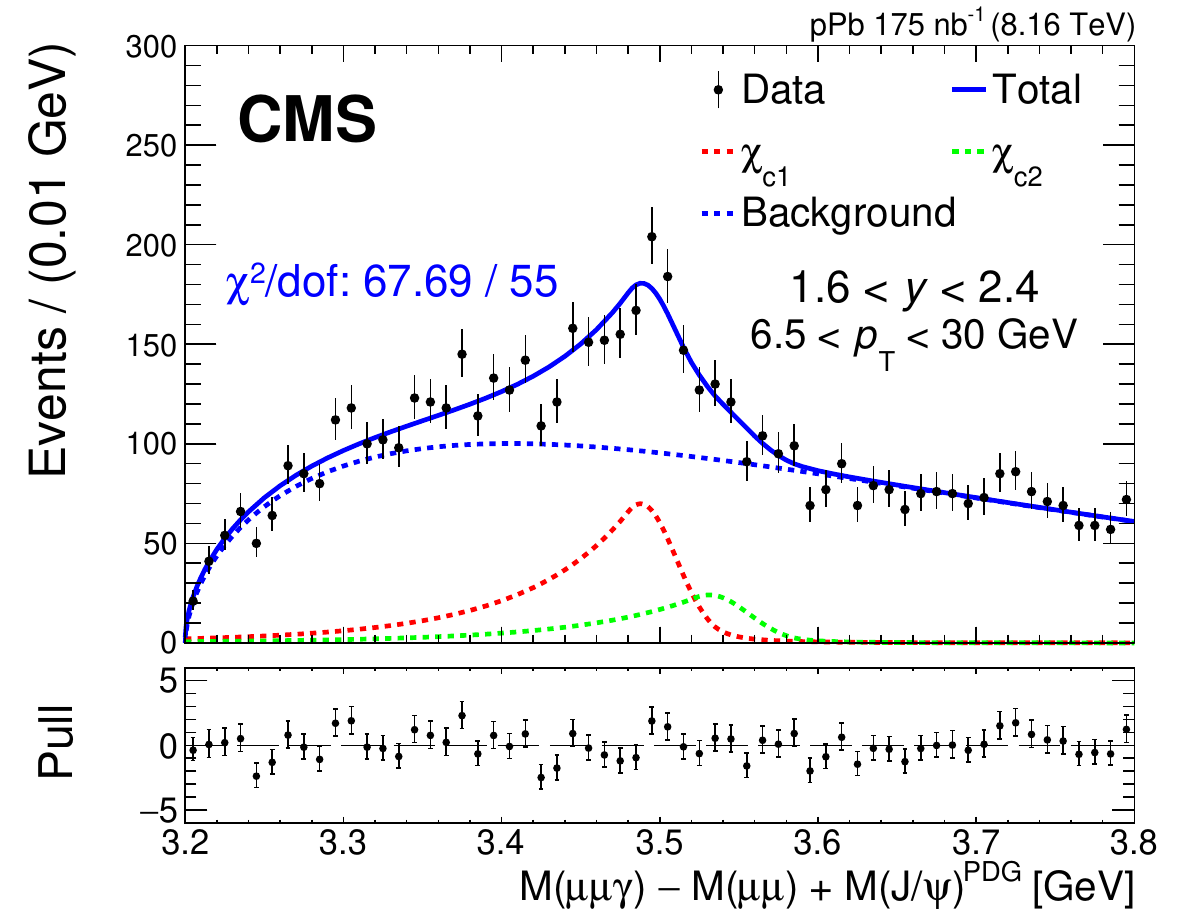}
  \caption{Examples of fits for \chic candidates, integrated over \pt(\JPsi) and \Ntracks. \cmsLLeft: midrapidity bin ($0.0<\ylabFlipJ<1.0$), \cmsRight: forward rapidity bin ($1.6<\ylabFlipJ<2.4$). The vertical lines of data points represent the statistical uncertainties. The overall fit is shown in blue; the dashed blue line is the background. The individual peak DCB functions are shown in red for \chicone and green for \chictwo. }
  \label{fig:ExampleFits}
\end{figure}

Figure~\ref{fig:ExampleFits} shows an example \chic candidate fit. 
The lower panel of Fig.~\ref{fig:ExampleFits} shows the bin-by-bin normalized residuals (pulls) between the data and the fit model. The goodness-of-fit measure $\chi^2$/dof is shown for each fit, where $\chi^2$ is the sum of the squares of the pulls for all bins, and dof is the number of degrees of freedom calculated as the number of bins minus the number of fit parameters.
In all cases, the fitted yield of the \chic states is nonzero, with observed significances above $5\sigma$.

\section{Systematic uncertainties}
\label{sec:Systematics}

One of the main advantages of studying the ratio of \chictwo-to-\chicone meson yields is that many of the systematic uncertainties are expected to partially cancel out, most notably those related to the integrated luminosity, \JPsi meson acceptance, and \JPsi reconstruction efficiency. The remaining uncertainties are discussed in this section.

\subsection{Uncertainties in signal fitting}
\label{sec:SignalFittingSyst}

The uncertainty from \chic candidate fitting is evaluated using the Gaussian-constraint method, which is applied to our original fitting results. This method penalizes the likelihood according to a Gaussian distribution, should the parameter deviate from the best fit value determined from the simulation analysis. The method also ensures the compatibility of the constraint fitting parameter without distorting its shape. The mean and standard deviation of the Gaussian constraint used for a given parameter is obtained from fits to the distributions in MC simulation, where the mean is the value of the fitted parameter and the standard deviation is its fit uncertainty. 
The uncertainty is calculated by the ratio of the nominal results to the Gaussian-constrained fit results.

The low event yields make statistical bias non-negligible. To account for this, we generate pseudo-data using the likelihood with known input parameters to study possible statistical biases. We calculate the mean bias of the pull distribution and multiply it by the relative statistical uncertainty to obtain the systematic uncertainty because of the statistical bias arising from the \chicone and \chictwo candidate fitting procedure. The lowest (highest) value is 1 (4)\%. These systematic uncertainties are also added to the final result.

\subsection{Uncertainties in the converted photon reconstruction}
\label{sec:SystematicsConversionSyst}

To obtain the systematic uncertainty from the correction factors for the photon conversion reconstruction, we use two approaches. The first approach is to apply an alternative conversion selection and examine the difference. The alternative selection is a different selection proposed during the process of revisiting and optimizing the existing CMS \chic analysis selection to fit the current analysis. In this context, because of the statistical effect, comparing the nominal result with a less restrictive alternative ---defined as the selection without applying the quality criteria described in Section~\ref{sec:Selection}--- is more reasonable than comparing it with a tighter selection. For each point, the relative differences of the results are recalculated using both the nominal and the less restrictive selection efficiencies and yields, and the differences are expressed as percentages. Then, for each kinematic variable of the analysis, these differences have their means and standard deviation values calculated. The uncertainty in each bin is obtained by adding the mean and the standard deviation in quadrature. This procedure provides a reasonable estimate of the potential variation in the results arising from the conversion selection. Since the variation of uncertainty for each kinematic variable is unlikely to be significant, the values from each set are averaged. As a result, a 6\% uncertainty is assigned to the \chictoJratio, and a 13\% uncertainty to the \chictwooneratio.

The second approach uses the \pizero meson decay process $ \pizero \to \PGg \PGg \to  \ee~+~\ee$ by which we can obtain the selection efficiency directly from data, thereby allowing verification of the MC efficiency. The denominator of the efficiency consists of the yield of \pizero candidates reconstructed using one conversion photon selected with very stringent selection criteria (tag) and one conversion photon selected with very loose criteria (probe). The yield used in the numerator consists of the number of \pizero candidates that remain after applying the analysis selections to the probe converted photon. We compared the differences between the efficiency derived from this data-driven method to that obtained from MC simulation. The efficiency difference is found to be 5\%  in the integrated bin, and this is assigned as the systematic uncertainty in the efficiency correction, estimated as the difference in the efficiency obtained from simulation to that obtained from data, \chictwooneratio.

\subsection{Uncertainties in correction factors derived from simulation}

The \chic states are reconstructed only if the photon was emitted at rapidity $\abs{y(\PGg)}<2.4$ and $\pt(\PGg) > 0.5$\GeV. We correct for the fraction of photons from the \chic meson decays that lie outside of the photon acceptance window. This correction is determined from the MC simulation and depends on the underlying assumptions of the \PYTHIA generator, namely on the \chic meson kinematic distribution it produces. As mentioned in Section~\ref{sec:DataMC}, the \chic meson distributions are constrained to data. This reduces the dependence of the results on the MC simulation settings and thus limits the uncertainty arising from the simulation details. The remaining effect is studied by comparing the acceptance and efficiency correction obtained from simulated samples that were generated with modified parameters of the \PYTHIA generator. The parameters that we vary are pTHatMin, the renormalization and factorization scales, and the c quark mass. The pTHatMin variable is minimum invariant \pt used for the generating parton interaction. The parameter pTHatMin is the minimum \pt of the partonic $2 \to 2$ process that generates the \chic meson, and plays a role in the shape of the final \pt distribution of the \chic states. The value of pTHatMin used for the nominal results is 4.5 \GeV, and was varied from 3.0 \GeV to 6.0 \GeV to estimate the uncertainty. The c-quark mass used for the nominal case is 1.5 \GeV, and is varied to 1.43 \GeV to estimate the uncertainty. Similarly, the nominal renormalization and factorization scales are taken to be 2-3 times the c-quark mass for uncertainty estimation. The maximum difference in the overall results is taken as the systematic uncertainty.

\subsection{Uncertainties in nonprompt \texorpdfstring{\JPsi}{J/psi} contamination}
Since nonprompt \JPsi meson production cannot be neglected at LHC energies, we estimate the contribution of nonprompt \JPsi mesons in the data set and assign it as a systematic uncertainty. It is found that the \chictwooneratio has a linear dependence on the non-prompt \JPsi meson contamination. We quantify the linear relationship by measuring both ratios using two different \JPsi meson lifetime selection criteria. We then assign the systematic uncertainty to be the difference between the ratios obtained with the nominal $c\tau$ selection and those estimated using the linear relationship evaluated at the point where the non-prompt \JPsi meson contamination would be zero. The overall uncertainties arising from the \JPsi meson contamination are found to be 8\%. 
\subsection{Overall systematic uncertainty}
\label{sec:Syst_overall}

Individual systematic uncertainties are summed in quadrature under the assumption that they are uncorrelated. Table~\ref{tab:OverallSystUncertainty} summarizes the sources of systematic uncertainty. For those uncertainties that depend on \ylabFlipJ, \ptJ, and \Ntracks, a range is given in the table. The dominant systematic uncertainties originate from the conversion selection and from the nonprompt contamination.

\begin{table}[hbtp]
\renewcommand{\arraystretch}{1.1} 
\centering
\topcaption{Overview of the systematic uncertainties.}
\cmsTable{
\begin{tabular}{lc}
Source of uncertainty & Uncertainty [\%] \\
\hline
\chic fit shape & 1--4 \\
Conversion selection & 13 \\
Conversion selection -- tag-and-probe & 5 \\
\PYTHIA settings & 0.1--0.5 \\
Nonprompt \JPsi meson contamination & 8 \\[0.5ex]
Total systematic uncertainty & 16--17 \\
\end{tabular}}
  
\label{tab:OverallSystUncertainty}
\end{table}

\section{Polarization effects}
\label{sec:Polarization}

The polarization of \chic states has been previously measured in \pp collisions~\cite{CMS:2019jas, Faccioli:2020rue}. These measurements strongly disfavor a scenario where both \chicone and \chictwo are unpolarized, and are instead compatible with \Jzerozero where $J_z$ is the projection of the \chic states total angular momentum along the z axis in the center of mass \JPsi rest frame (``helicity'' frame). In contrast, the existing \JPsi meson polarization measurements are consistent with no polarization at low-\pt ~\cite{CMS:2013gbz, ALICE:2018crw}. Therefore, the results presented in this paper assume \Jzerozero, and unpolarized \JPsi. 

The polarization was taken into account by modifying the dimuon angular decay distributions in the rest frame of the \JPsi from \chic meson decay, following the procedure of Ref.~\cite{Faccioli:2011be}. The decay distribution of unpolarized \chic mesons in the simulated samples is flat in $\cos\theta$, where $\theta$ is the angle between  the momentum vector of the positively charged muon and that of the \JPsi meson in its rest frame at the generator-level. The MC angular distribution is weighted such that \ddinline{N}{(\cos \theta)} $ \propto 1 + \lambda_{\theta} \cos^{2}\theta$, where $\lambda_{\theta}$ is the parameter describing the polarization. The \Jzerozero polarization condition corresponds to $\lambda_{\theta}=1$ for \chicone and $-0.6$ for \chictwo.  The results are corrected for acceptance and efficiency using the re-weighted \ddinline{N}{(\cos \theta)} distributions.  A positive (negative) $\lambda_{\theta}$ decreases (increases) the acceptance which is the dominant correction factor.  Because the $\lambda_{\theta}$ parameters have opposite signs in this scenario, the modified polarization has a noticeable effect on the \chictwooneratio result. The size of this effect is illustrated in the next section by comparing the polarized and unpolarized results.

\section{Results and discussion}
\label{sec:Results}

The \chictwooneratio was determined as a function of \Ntracks, \ylabFlipJ, and \ptJ to probe potential differential modification in the nuclear medium.

\begin{figure}[hb!t]
  \centering
	\includegraphics[width=0.48\textwidth]{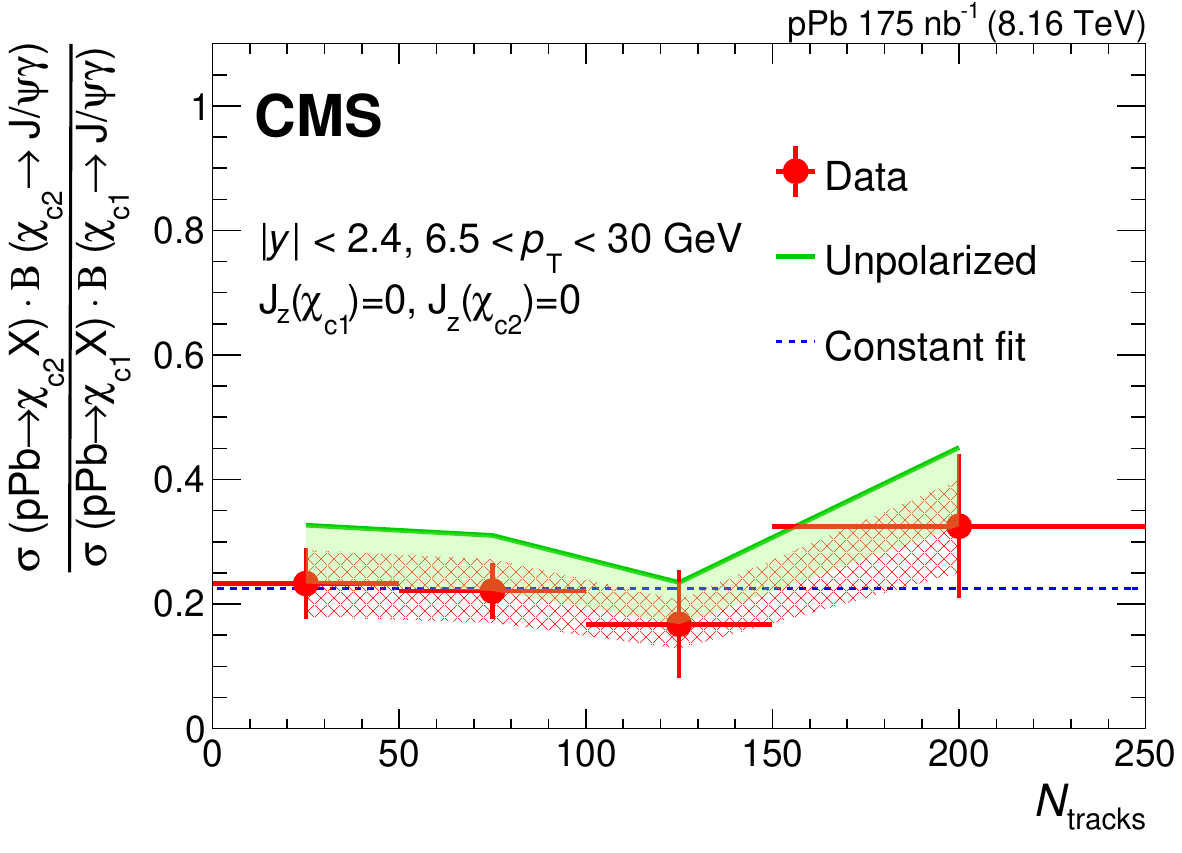}
  \includegraphics[width=0.48\textwidth]{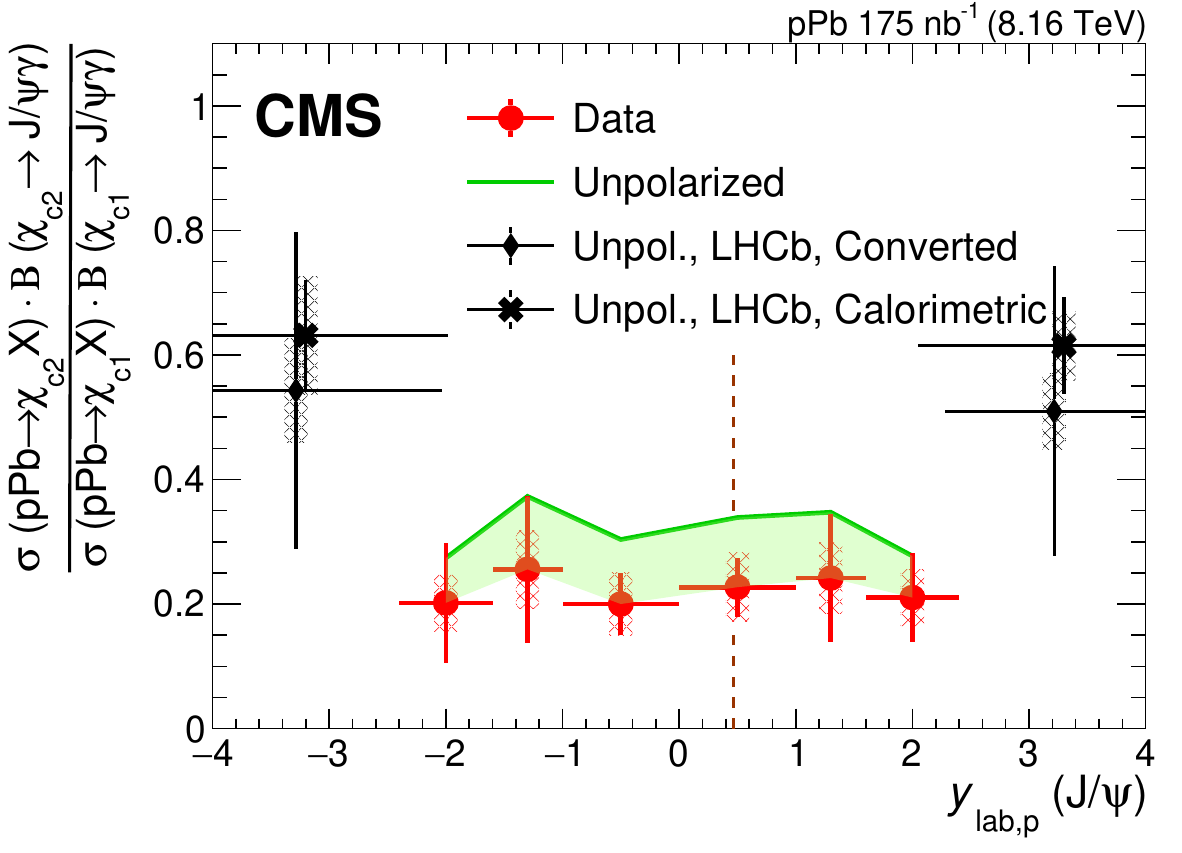}
  \caption{The \chictwooneratio as a function of \Ntracks (\cmsLeft) and as a function of rapidity \ylabFlip (\cmsRight). The red points are for the \Jzerozero scenario, while the green line represents the unpolarized case. The light green band indicates the difference between the results obtained under the unpolarized and polarization hypotheses. The hashed boxes on the points represent the systematic uncertainties. The blue dashed line shows the result of a constant fit to the points in the \cmsLeft panel. The rapidity is defined in the lab frame, with the p-going direction being positive. In the right panel, the center-of-mass rapidity is shown by the brown dashed vertical line at $\ylab = 0.465$. Measurements reported by LHCb are shown in black~\cite{bib:chictochicpPbLHCb}.}
  \label{fig:Results_nTrk}
\end{figure}

The \cmsLeft panel in Fig.~\ref{fig:Results_nTrk} presents this ratio as a function of event multiplicity (\Ntracks).
The red points represent the results for the \Jzerozero scenario, while the green line indicates the results assuming no \chic polarization.
A constant function fit to the data yields $C= 0.23 \pm 0.03$ ($\chi^2/\mathrm{dof} = 1.2/3$, $p$-value $= 0.74$).
The observed absence of prominent multiplicity dependence suggests that both \chic states are influenced by the nuclear medium in a similar manner.
This behavior stands in contrast to the $\PGyP{2S}$-to-$\JPsi$ sector, where a significant relative suppression of the excited state is typically observed in high-multiplicity collisions.

To further test for possible kinematic-dependent differences, the results are examined as a function of rapidity in the \cmsRight panel of Fig.~\ref{fig:Results_nTrk}. 
The center-of-mass rapidity of the proton-nucleon system is indicated by the brown dashed line. 
Similar to the multiplicity dependence, the \chictwooneratio remains constant within uncertainties across the full rapidity range. 
This result is compared with LHCb measurements from \pPb collisions~\cite{bib:chictochicpPbLHCb}, which provide an adjacent rapidity coverage. For a proper comparison, the LHCb results are corrected for the corresponding branching ratios.
Since the LHCb measurement was performed under a no-polarization assumption, we compare it to the corresponding green line from this study, and observe no strong indications for rapidity-dependent modification, given the uncertainties.

\begin{figure}[hb!t]
  \centering
  \includegraphics[width=0.48\textwidth]{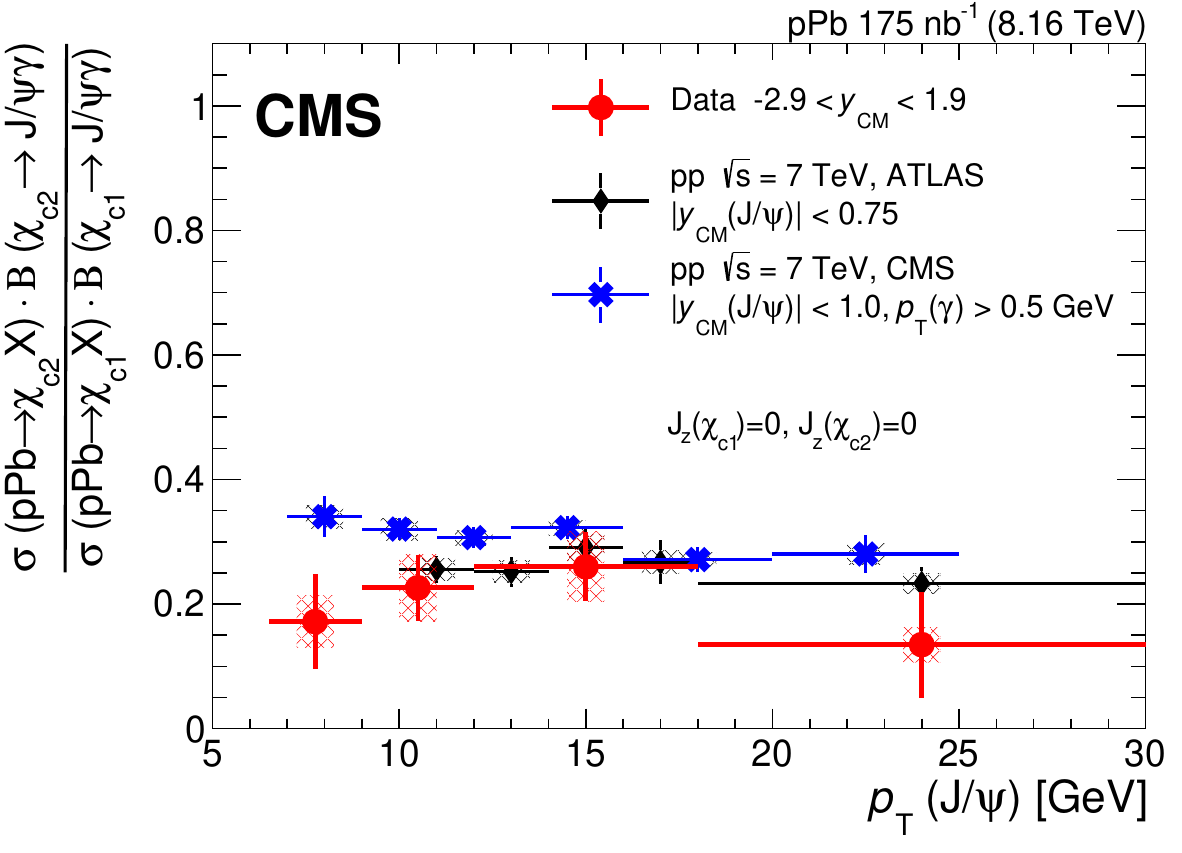}
  \caption{The \chictwooneratio as a function of \ptJ, shown by red points for the \Jzerozero  polarization assumption. The results are integrated over all the detector acceptance $\abs{\ylab(\JPsi)}<2.4$, corresponding to $-2.9<\rap<1.9$, and over the full charged-particle multiplicity range. The hashed boxes on the points represent the systematic uncertainties. The \chictwooneratio is compared with existing \pp measurements by ATLAS~\cite{bib:chicATLAS2014} (black diamonds) and CMS~\cite{bib:chictochicratioCMS2012} (blue crosses) with the same polarization assumption.}
  \label{fig:Results_pt}
\end{figure}

Figure~\ref{fig:Results_pt} displays the \ptJ dependence of the \chictwooneratio. 
The results exhibit no significant variation with \ptJ ($C= 0.22 \pm 0.05$, $\chi^2/\mathrm{dof} = 1.9/3$, $p$-value $= 0.59$). The results are compared with existing pp midrapidity measurements from ATLAS ($\abs{y}<0.75$)~\cite{bib:chicATLAS2014} and CMS ($\abs{y}<0.75$)~\cite{bib:chictochicratioCMS2012}. Overall, the pPb results are compatible with the pp measurements within uncertainties, although some differences between the CMS pp results are visible in the overlapping \pt range, particularly at low \pt. 

This stability in the production ratio of P-wave states stands in contrast to the behavior of S-wave charmonium states. 
The previous CMS studies have shown a $5.3\sigma$ significant multiplicity-dependent suppression of the $\PGyP{2S}/\JPsi$ ratio~\cite{CMS:2025oqe}, indicating that the pPb system creates a dense environment capable of dissociating weakly bound states. 
While the nuclear medium effectively resolves the large size difference between \PGyP{2S} (0.45\unit{fm}) and $\JPsi$ (0.25\unit{fm}), leading to sequential suppression, it fails to differentiate between $\chicone$ and $\chictwo$~\cite{Satz:2006kba}.
These two P-wave states share nearly identical radii ($\sim$0.39\unit{fm}) despite their mass splitting ($\Delta M \approx 45\MeV$). 
The constant \chictwooneratio implies that the dissociation mechanism is primarily driven by the macroscopic geometric size of the quarkonium state rather than its specific spin configuration or fine-structure binding energy.

Furthermore, the consistency between pPb and \pp results is compatible with the expectation that no sizable relative modification between the \chicone and \chictwo states should be observed within the current uncertainties, given their similar radii and small fine-structure mass splitting ($E_b(\chicone) - E_b(\chictwo) \approx 45\MeV$). 
The present measurement therefore shows no evidence for a differential modification of the two \(P\)-wave charmonium states in pPb collisions.
This observation is consistent with a picture in which medium effects are similar for states with comparable spatial sizes.

The interpretation of these findings requires a careful consideration of systematic uncertainties. 
While the $\PGyP{2S}/\JPsi$ analysis benefited from the cancellation of most systematic errors (yielding 1--4\% precision), the present $\chictwo/\chicone$ analysis carries larger uncertainties (16--17\%), dominated by photon conversion efficiency. 
However, the central values consistently show a flat distribution across multiplicity, \ptJ, and rapidity, and this trend remains robust under various polarization scenarios. 
Therefore, despite the larger statistical and systematic limitations, the conclusion regarding the absence of relative modification between P-wave states remains solid.

\section{Summary}
\label{sec:Summary}

The measurement of the ratio $[\sigma (\chictwo) \mathcal{B}(\chictwo\to\JPsi\PGg)] / [\sigma (\chicone) \mathcal{B}(\chicone\to\JPsi\PGg)] $ was presented in \pPb collisions at \sqrtsNNeight. The results are reported as a function of event charged-particle multiplicity \Ntracks, rapidity \ylabFlipJ, and transverse momentum \ptJ of the \JPsi. The \chicone and \chictwo states were reconstructed through the decay \chictoJgfull .  The measurements were reported in the CMS detector acceptance region $6.5<\pt(\JPsi)<30\GeV$ and $\abs{\ylabFlip(\JPsi)}<2.4$, corresponding approximately to $-2.9<\rap(\JPsi)<1.9$ in the proton-nucleon center-of-mass system.

No significant dependence of the \chictwooneratio on \Ntracks, \ptJ, or \ylabFlipJ was observed, and the results are consistent with existing \pp measurements. 
These findings demonstrate that the relative production of these P-wave states remains unmodified in \pPb collisions.
This absence of a significant multiplicity dependence stands in contrast to the S-wave sector, where \PGyP{2S} suppression has been interpreted as evidence for final-state effects in pPb collisions. 
The multiplicity independence of the \chic ratio suggests that any dissociation or final-state modification mechanism does not produce a measurable difference between the \chicone and \chictwo states within the current uncertainties.
The \pPb medium appears capable of resolving the large size disparities between S-wave states, leading to sequential suppression, while for the P-wave states \chicone and \chictwo, the results do not indicate a clear sensitivity to their more subtle internal differences.

\begin{acknowledgments}
We congratulate our colleagues in the CERN accelerator departments for the excellent performance of the LHC and thank the technical and administrative staffs at CERN and at other CMS institutes for their contributions to the success of the CMS effort. In addition, we gratefully acknowledge the computing centers and personnel of the Worldwide LHC Computing Grid and other centers for delivering so effectively the computing infrastructure essential to our analyses. Finally, we acknowledge the enduring support for the construction and operation of the LHC, the CMS detector, and the supporting computing infrastructure provided by the following funding agencies: SC (Armenia), BMBWF and FWF (Austria); FNRS and FWO (Belgium); CNPq, CAPES, FAPERJ, FAPERGS, and FAPESP (Brazil); MES and BNSF (Bulgaria); CERN; CAS, MoST, and NSFC (China); MINCIENCIAS (Colombia); MSES and CSF (Croatia); RIF (Cyprus); SENESCYT (Ecuador); ERC PRG and PSG, TARISTU24-TK10 and MoER TK202 (Estonia); Academy of Finland, MEC, and HIP (Finland); CEA and CNRS/IN2P3 (France); SRNSF (Georgia); BMFTR, DFG, and HGF (Germany); GSRI (Greece); MATE and NKFIH (Hungary); DAE and DST (India); IPM (Iran); SFI (Ireland); INFN (Italy); MSIT and NRF (Republic of Korea); MES (Latvia); LMTLT (Lithuania); MOE and UM (Malaysia); BUAP, CINVESTAV, CONACYT, LNS, SEP, and UASLP-FAI (Mexico); MOS (Montenegro); MBIE (New Zealand); PAEC (Pakistan); MSHE, NSC, and NAWA (Poland); FCT (Portugal); MESTD (Serbia); MICIU/AEI and PCTI (Spain); MOSTR (Sri Lanka); Swiss Funding Agencies (Switzerland); MST (Taipei); MHESI (Thailand); TUBITAK and TENMAK (T\"{u}rkiye); NASU (Ukraine); STFC (United Kingdom); DOE and NSF (USA).

\begin{sloppypar}
\setlength\emergencystretch{\hsize}
\hyphenation{Rachada-pisek} Individuals have received support from the Marie-Curie program and the European Research Council and Horizon 2020 Grant, contract Nos.\ 675440, 724704, 752730, 758316, 765710, 824093, 101115353, 101002207, 101001205, and COST Action CA16108 (European Union); the Leventis Foundation; the Alfred P.\ Sloan Foundation; the Alexander von Humboldt Foundation; the Science Committee, project no. 22rl-037 (Armenia); the Fonds pour la Formation \`a la Recherche dans l'Industrie et dans l'Agriculture (FRIA) and Fonds voor Wetenschappelijk Onderzoek contract No. 1228724N (Belgium); the Beijing Municipal Science \& Technology Commission, No. Z191100007219010, the Fundamental Research Funds for the Central Universities, the Ministry of Science and Technology of China under Grant No. 2023YFA1605804, the Natural Science Foundation of China under Grant No. 12535004, and USTC Research Funds of the Double First-Class Initiative No.\ YD2030002017 (China); the Ministry of Education, Youth and Sports (MEYS) of the Czech Republic; the Shota Rustaveli National Science Foundation (Georgia); the Deutsche Forschungsgemeinschaft (DFG), among others, under Germany's Excellence Strategy -- EXC 2121 ``Quantum Universe" -- 390833306, and under project number 400140256 - GRK2497; the Hellenic Foundation for Research and Innovation (HFRI), Project Number 2288 (Greece); the Hungarian Academy of Sciences, the New National Excellence Program - \'UNKP, the NKFIH research grants K 131991, K 138136, K 143460, K 143477, K 147557, K 146913, K 146914, K 147048, TKP2021-NKTA-64, and 2025-1.1.5-NEMZ\_KI-2025-00004, and MATE KKP and KKPCs Research Excellence and Flagship Research Groups grants (Hungary); the Council of Science and Industrial Research, India; ICSC -- National Research Center for High Performance Computing, Big Data and Quantum Computing, FAIR -- Future Artificial Intelligence Research, and CUP I53D23001070006 (Mission 4 Component 1), funded by the NextGenerationEU program, the Italian Ministry of University and Research (MUR) under Bando PRIN 2022 -- CUP I53C24002390006, PRIN PRIMULA 2022RBYK7T (Italy); the Latvian Council of Science; the Ministry of Science and Higher Education, project no. 2022/WK/14, and the National Science Center, contracts Opus 2021/41/B/ST2/01369, 2021/43/B/ST2/01552, 2023/49/B/ST2/03273, and the NAWA contract BPN/PPO/2021/1/00011 (Poland); the Funda\c{c}\~ao para a Ci\^encia e a Tecnologia (Portugal); the National Priorities Research Program by Qatar National Research Fund; MICIU/AEI/10.13039/501100011033, ERDF/EU, ``European Union NextGenerationEU/PRTR", projects PID2022-142604OB-C21, PID2022-139519OB-C21, PID2023-147706NB-I00, PID2023-148896NB-I00, PID2023-146983NB-I00, PID2023-147115NB-I00, PID2023-148418NB-C41, PID2023-148418NB-C42, PID2023-148418NB-C43, PID2023-148418NB-C44, PID2024-158190NB-C22, RYC2021-033305-I, RYC2024-048719-I, CNS2023-144781, CNS2024-154769 and Plan de Ciencia, Tecnolog{\'i}a e Innovaci{\'o}n de Asturias, Spain; the Chulalongkorn Academic into Its 2nd Century Project Advancement Project, the National Science, Research and Innovation Fund program IND\_FF\_68\_369\_2300\_097, and the Program Management Unit for Human Resources \& Institutional Development, Research and Innovation, grant B39G680009 (Thailand); the Eric \& Wendy Schmidt Fund for Strategic Innovation through the CERN Next Generation Triggers project under grant agreement number SIF-2023-004; the Kavli Foundation; the Nvidia Corporation; the SuperMicro Corporation; the Welch Foundation, contract C-1845; and the Weston Havens Foundation (USA).
\end{sloppypar}

\end{acknowledgments}\section*{Data availability} Release and preservation of data used by the CMS Collaboration as the basis for publications is guided by the  \href{https://doi.org/10.7483/OPENDATA.CMS.1BNU.8V1W}{CMS data preservation, re-use and open access policy}.

\bibliography{auto_generated}   

\cleardoublepage \appendix\section{The CMS Collaboration \label{app:collab}}\begin{sloppypar}\hyphenpenalty=5000\widowpenalty=500\clubpenalty=5000\input{HIN-22-003-public-authorlist.tex}\end{sloppypar}
\end{document}

%% file: HIN-22-003-public-authorlist.tex
\cmsinstitute{Yerevan Physics Institute, Yerevan, Armenia}
{\tolerance=6000
A.~Belyaev\cmsorcid{0000-0003-1692-1173}, A.~Hayrapetyan, A.~Tumasyan\cmsAuthorMark{1}\cmsorcid{0009-0000-0684-6742}
\par}
\cmsinstitute{Institut f\"{u}r Hochenergiephysik, Vienna, Austria}
{\tolerance=6000
W.~Adam\cmsorcid{0000-0001-9099-4341}, J.W.~Andrejkovic, L.~Benato\cmsorcid{0000-0001-5135-7489}, T.~Bergauer\cmsorcid{0000-0002-5786-0293}, M.~Dragicevic\cmsorcid{0000-0003-1967-6783}, C.~Giordano\cmsorcid{0000-0001-6317-2481}, P.S.~Hussain\cmsorcid{0000-0002-4825-5278}, M.~Jeitler\cmsAuthorMark{2}\cmsorcid{0000-0002-5141-9560}, N.~Krammer\cmsorcid{0000-0002-0548-0985}, A.~Li\cmsorcid{0000-0002-4547-116X}, D.~Liko\cmsorcid{0000-0002-3380-473X}, M.~Matthewman, I.~Mikulec\cmsorcid{0000-0003-0385-2746}, J.~Schieck\cmsAuthorMark{2}\cmsorcid{0000-0002-1058-8093}, D.~Schwarz\cmsorcid{0000-0002-3821-7331}, R.~Sch\"{o}fbeck\cmsAuthorMark{2}\cmsorcid{0000-0002-2332-8784}, M.~Shooshtari\cmsorcid{0009-0004-8882-4887}, M.~Sonawane\cmsorcid{0000-0003-0510-7010}, W.~Waltenberger\cmsorcid{0000-0002-6215-7228}, C.E.~Wulz\cmsAuthorMark{2}\cmsorcid{0000-0001-9226-5812}
% ...
\par}
$^{1}$Also at Yerevan State University, Yerevan, Armenia\\
$^{2}$Also at Technische Universit\"{a}t Wien, Vienna, Austria\\
% ...
